\documentclass[%
 reprint,
 amsmath,amssymb,
 aps,
]{revtex4-1}

\usepackage{graphicx}
\usepackage{dcolumn}
\usepackage{bm}

\begin{document}

\preprint{APS/123-QED}


\title{Nodal Superconducting Gap in $\beta$-FeS}

\author{Jie Xing, Hai Lin, Yufeng Li, Sheng Li, Xiyu Zhu, Huan Yang and Hai-Hu
Wen$^{\star}$}

\affiliation{National Laboratory of Solid State Microstructures and Department of Physics,
Collaborative Innovation Center of Advanced Microstructures, Nanjing University, Nanjing 210093, China}

\begin{abstract}
Low temperature specific heat has been measured in superconductor $\beta$-FeS with T$_c$ = 4.55 K. It is found that the low temperature electronic specific heat C$_e$/T can be fitted to a linear relation in the low temperature region, but fails to be described by an exponential relation as expected by an s-wave gap. We try fittings to the data with different gap structures and find that a model with one or two nodal gaps can fit the data. Under a magnetic field, the field induced specific heat coefficient $\Delta\gamma_e$=[C$_e$(H)-C$_e$(0)]/T shows the Volovik relation $\Delta\gamma_e(H)\propto \sqrt{H}$, suggesting the presence of nodal gap(s) in this material.
\begin{description}
\item[Subject Areas]
Condensed Matter Physics, Superconductivity
\end{description}
\end{abstract}
\maketitle

\section{\label{sec:level1}INTRODUCTION}
The discovery of superconductivity in iron pnictides has opened a new era for the study on
unconventional superconductors\cite{Kamihara2008}. Until now, superconductivity has been found in several relatives of the FeAs-based superconductors, including the FeSe-based families and the iron phosphides\cite{stewart,RGreene,review}. In either FeAs-based or FeSe-based compounds, the superconducting transition temperature can be raised to much higher than 40 K\cite{REN,llsun}, the so-called McMillan limit\cite{mcmillan}. This allows us to categorize the iron based superconductors as an unconventional superconductor with the pairing mechanism beyond the electron-phonon coupling. In the research on the iron based superconductors, one of the core issues is about the superconducting pairing mechanism. Early on, theoretical pictures concerning the s$^\pm$ pairing gap was proposed\cite{MazinPRL,KurokiPRL}, this model has got many experimental supports\cite{HanaguriQPI,Neutron,CuImpurity} and is specially satisfied by the systems with the presence of both electron and hole pockets. Recently, superconductivity has been reported\cite{LinJH} in $\beta$-FeS which has an isostructure of the $\beta$-FeSe superconductor\cite{WuMK}. Earlier band structure calculations have revealed that the electronic structures of the FeSe and FeS systems are quite similar to each other\cite{Singh}. Usually the $p$-orbital of the element sulfur has a narrower bandwidth comparing with that of selenium. Therefore, it is interesting to know the superconducting gap structure and pairing mechanism in this newly discovered superconductor.

Superconductivity is induced by the condensation of Cooper pairs formed by electron-boson interactions (or strong local magnetic superexchange), such as electron-phonon coupling in the Bardeen-Copper-Schrieffer (BCS) theory. The superconducting gap $\Delta(k)$ that protects the superconducting condensate from exciting the quasiparticles is momentum dependent and is intimately related to the underlying pairing interactions. A $d$-wave gap has been observed in cuprates in revealing the unconventional mechanism\cite{Tsuei}. While accidental nodes are sometimes observed in iron pnictide superconductors reflecting a basic feature of the extended s-wave\cite{MazinPRL,KurokiPRL,hirschfeldReview}. At the gap nodes, the magnitude of a superconducting gap is zero and the signs are opposite at two sides of the nodal point. The nodal gaps, if enforced by symmetry, would suggest a pairing mechanism with repulsive interactions, such as that in the cuprates. Therefore to determine the gap structures and know whether the gap is nodal or nodeless is an effective method to check or even prove the mechanism of superconductivity. Theoretically it is understood that the low-energy quasi-particle (QP) excitations for nodal gap superconductors is much stronger than that of the convention superconductors with s-wave gaps and the magnetic field induced generation of QP density of states (DOS) has very different field dependence. Since the discovery of iron-based superconductors, nodal gap(s) were suggested or indicated in several superconducting systems, such as LaFePO\cite{LaFePO}, BaFe$_2$As$_{2-x}$P$_x$\cite{BaFeAsP}, and KFe$_2$As$_2$\cite{KFeAs1,KFeAs2,KFeAs3}. Although early works about the specific heat and thermal conductivity in FeSe suggested multiple nodeless gaps\cite{FeSeSH,FeSeHT}, the recent STM works on FeSe single crystal and FeSe film supported that there are line nodes in those samples\cite{QKXue,Hanaguri}.

\begin{figure}
\includegraphics[width=8cm]{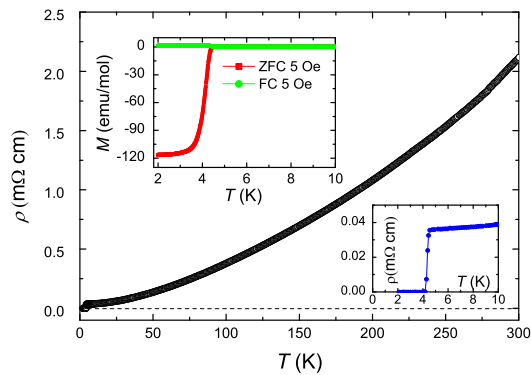}
\caption {(color online) Temperature dependence of resistivity at zero magnetic field. Upper left inset: Temperature dependence of
magnetic susceptibility measured in the zero-field-cooled mode (ZFC) and the field-cooled mode (FC) at H = 5Oe. Lower right inset: The enlarged view
of resistivity for FeS from 2K to 10K.}
\label{fig1}
\end{figure}

In this paper, we report the investigations on low temperature specific heat in $\beta$-FeS crystals. A linear relationship between C$_e$/T and T is observed in the low temperature limit, this is deviating from the exponential relation expected for s-wave gaps and indicates nodal superconducting gap(s). It is further found that the fitting to the electronic specific heat by a model with two nodal gaps is better than that by adopting either a single s- or d-wave gaps or a mixture of s-wave and d-wave gaps. Furthermore, magnetic field induced electronic specific heat shows consistency with the Volovik relation $\Delta\gamma_e(H)\propto \sqrt{H}$ approximately\cite{volovik}, which was predicted specially for $d$-wave superconductors with line nodes. The analysis of low temperature specific heat reveals the presence of nodal superconducting gap(s) in this system, which is in accordance with the recent heat transport experiment\cite{FeSHT}.

\section{\label{sec:level2}EXPERIMENTAL METHOD}

The FeS crystals were fabricated by a hydrothermal method which has been presented before\cite{HLin}. The resistivity and low temperature specific heat were measured with a Quantum Design instrument PPMS-16T, and the magnetization was detected by a Quantum Design instrument SQUID-VSM. The specific heat measurements were done with the thermal relaxation method by He3 accessory of PPMS-16T with an advanced single crystal quatz platform which has a quite weak magnetic field effect. All the data in this paper were measured on samples from the same batch FeS crystals.

\section{\label{sec:level2}RESULTS AND DISCUSSION}

Fig.~1 shows the temperature dependence of resistivity at zero field. As can be seen in the lower right inset, a sharp resistive transition occurs with the onset transition temperature of 4.55 K, and the zero resistivity appears at about 4.15 K. The normal state resistivity shows a monotonous decreasing with temperature from 300K to 4.55K, which indicates metallic conductivity. The residual resistivity ratio (RRR) defined as $\rho(300K)/\rho(0K)$ is about 60, indicating a good quality of the sample. This result is similar to the former reported value\cite{HLin}. The upper left inset shows the temperature dependence of magnetic susceptibility measured in the zero-field-cooled (ZFC) and the field-cooled (FC) mode. A rough estimate on the magnetic signal yields a full superconducting screening volume. At an external magnetic field of 5Oe the diamagnetic transition starts near 4.5K, which is accordance with the resistivity measurement very well.

\begin{figure}
\includegraphics[width=8cm]{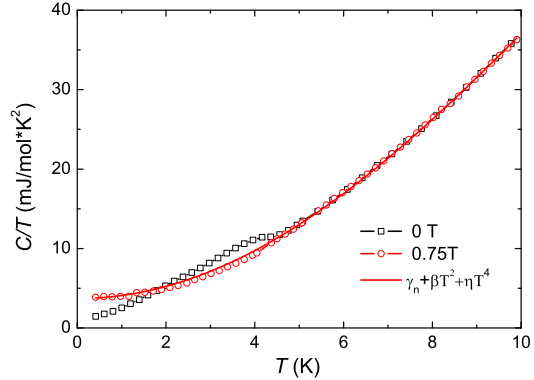}
\caption {(color online) Raw data of the temperature dependence of
specific heat for FeS crystals at 0T (open square) and 0.75T (open circles). The red solid line is the normal state fitting curve by using the formula $C/T=\gamma_n+\beta T^2+\eta T^4$.} \label{fig2}
\end{figure}

The raw data of specific heat at 0T and 0.75T are shown in Fig.~2. An obvious superconducting anomaly appears at about 4.5K, which is also consistent with the resistive and magnetization data. The residual specific heat coefficient $\gamma_0$ at 0K and zero external magnetic field is about $1.1mJ/mol K^2$. Although this value is rather low, but comparing with the normal state Sommerfeld constant $\gamma_n$, it is quite sizable. As the magnetic field is increased to 0.75T, the superconducting anomaly is already suppressed to below 0.4K and the extracted $\gamma_n$ is about 3.8$\pm$0.2 $mJ/molK^2$. We try to fit the normal state specific heat data with the Debye model in low temperature region, however, the results hardly match the data well if using only the term $C_{ph}/T=\beta T^2$. We thus use the following model that contains three terms to fit the data from 0.4K to 15K. The fitting formula reads
\begin{equation}
C/T=\gamma_n +\beta T^2 + \eta T^4.
\end{equation}
The fitting curve as shown by the solid line in Fig.2 is better than the simple model and yields parameters $\gamma_n=3.8mJ/mol K^2$, $\beta=0.39 mJ/mol K^4$, and $\eta=-0.00053 mJ/mol K^6$. Using the relation between Debye temperature $\Theta_D$ and the fitting parameter $\beta$, $\Theta_D=(12\pi^4k_BN_AZ/5\beta)^{1/3}$, where $N_A$ = 6.02 $\times 10^{23}$
mol$^{-1}$ is the Avogadro constant, Z = 2 is the number of atoms in
one unit cell, we can calculate the Debye temperature of FeS, which is about 215K. The Debye temperature derived here for FeS is close to $\Theta_D$ = 210K obtained previously for FeSe single crystals\cite{FeSeSH}. The $\gamma_n$ seems to be smaller than that of other iron-based superconductors, for example, in FeSe single crystals $\gamma_n$ = 5.73$\pm$0.19 $mJ/mol K^2$\cite{FeSeSH}. According to the density functional calculations\cite{Singh}, it is found that the DOS in FeSe and FeS are quite close to each other. The smaller $\gamma_n$ finds here may suggest weaker electron-boson coupling or correlation effect in FeS. We present the specific heat under different magnetic fields in Fig.~3(a). By applying several magnetic fields up to 0.75T, the superconducting anomaly shifts to lower temperatures gradually and is hardly observed above 0.4K when the magnetic field is over 0.75T.
\begin{figure}
\includegraphics[width=8cm]{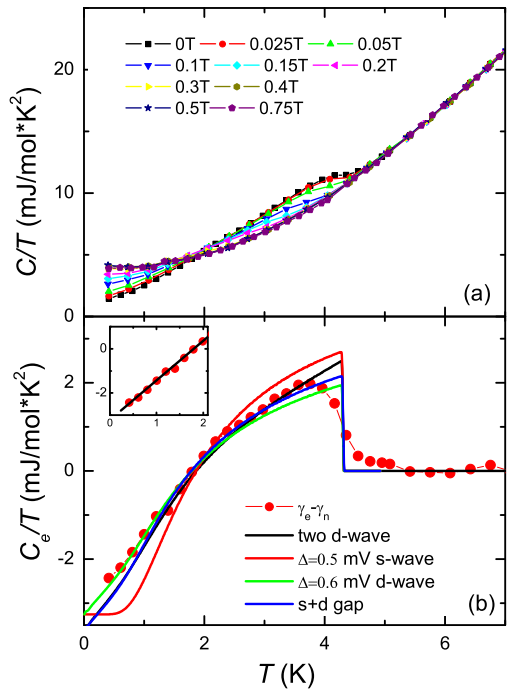}
\caption {(color online) (a)Temperature dependence of specific heat at different magnetic fields ranging from 0T to 0.75T. (b) The electronic specific heat of FeS crystals. The solid lines represent the fitting curves using the BCS formula by assuming four different gap(s): (1) single s-wave gap $\Delta(T,\theta)=\Delta_0(T)$; (2) single anisotropic gap with d-wave feature $\Delta(T,\theta)=\Delta_0(T)cos2\theta$; (3) mixture of two components with one s-wave and one d-wave; (4) mixture of two d-wave gaps. The inset shows the electronic specific heat coefficient $C_e$/T vs. T below 2K. The solid line shows a fit to the data and indicates a linear relation between $C_e$/T and T.} \label{fig3}
\end{figure}

The superconducting electronic contribution can be obtained by deducting the phonon contribution $C_{ph}$ below $T_c$. From Fig.~3 it is clear that the normal state specific heat does not depend on the magnetic field. Since a weak magnetic field 0.75T is enough to suppress the superconductivity in FeS, we can obtain the electronic specific heat through formula C$_e$=C(0T)-C(0.75T). The electronic specific coefficient C$_e$ derived in this way is shown in Fig.~3(b). It is found that the superconducting anomaly is not high, although the transition is rather sharp. Considering the entropy conservation near the superconducting transition, we determined the superconducting transition temperature T$_c$ = 4.3K from the the specific heat, and the jump $\Delta C/T_c$ is $2.47 mJ/mol K^2$. This leads to the ratio $\Delta C/\gamma_nT_c \approx$ 0.65. The value is much lower than the predicted one by the BCS theory in the weak coupling regime which gives $\Delta C/\gamma_nT_c$=1.43. This is also rather lower than 1.65 obtained for FeSe single crystals\cite{FeSeSH}. The specific heat anomaly in FeS is much lower than that in KFe$_2$As$_2$ although the transition temperature are close to each other\cite{Hardy}. This value of $\Delta C/\gamma_n T_c\mid_{Tc}$ reveals that FeS system may be a superconductor with rather weak electron-boson coupling and correlation effect. In KFe$_2$As$_2$ the Hund's rule coupling becomes stronger, this is assumed to be the origin of the correlation effect\cite{Hardy}. One can see that a tail appears in the specific heat data above T$_c$, which may be caused by the superconducting fluctuation. The inset of Fig.~3(b) shows the enlarge view of the low temperature specific heat data by fitting to a straight line. Interestingly, the electronic specific heat coefficient $C_e/T$ varies linearly with temperature in the low temperature regime. This relation is deviated from an exponential dependence as expected for an s-wave gap and can be explained by models containing nodal gaps. To investigate the details concerning superconducting gap(s), we use the BCS formula to fit our electronic specific heat data in the superconducting state

\begin{eqnarray}
\gamma_\mathrm{e}=\frac{4N(E_F)}{k_BT^{3}}\int_{0}^{+\infty}\int_0^{2\pi}\frac{e^{\zeta/k_BT}}{(1+e^{\zeta/k_BT})^{2}}\nonumber\\
\nonumber\\
(\varepsilon^{2}+\Delta^{2}(\theta,T)-\frac{T}{2}\frac{d\Delta^{2}(\theta,T)}{dT})\,d\theta\,d\varepsilon,
\end{eqnarray}

where $\zeta=\sqrt{\varepsilon^2+\Delta^2(T,\theta)}$. We try to use four different gap structures to fit the electronic specific heat data: (1) single s-wave gap $\Delta(T,\theta)=\Delta_0(T)$; (2) single anisotropic gap with d-wave feature $\Delta(T,\theta)=\Delta_0(T)cos2\theta$; (3) mixture of two components with one s-wave and one d-wave gap; (4) mixture of two d-wave gaps. If more than one gap is concerned in the calculation, a linear combination of the two components is used to get the electronic specific heat, namely $\gamma_e=p\gamma_e(\Delta_1)+(1-p)\gamma_e(\Delta_2)$ with $p$ the fraction contributed by the component of $\Delta_1$. The fitting parameters are listed in table 1. In Fig.~3(b) we can see that a simple s-wave or d-wave can not fit the data, although the fitting with a single d-wave gap is better. A model with the mixture of two d-wave gaps, as shown by the black solid line, gives the best fit, albeit a slight deviation in the low temperature region. The model with a mixture of s-wave and d-wave, but with large fraction of d-wave component could also describe the data roughly. The detailed fitting reveals that there should be anisotropic gaps with nodes in the FeS system. From previous studies in FeSe\cite{FeSeSH}, it was shown that the electronic specific heat could be described by a model with one anisotropic s-wave component, but with no sign of nodes in FeSe. However, recent scanning tunneling spectroscopy (STS) measurements on FeSe crystals\cite{Hanaguri} or thick film\cite{QKXue} show that line nodes exist in this system. These different gap structures of FeSe need more time to be resolved.
\begin{figure}
\includegraphics[width=8cm]{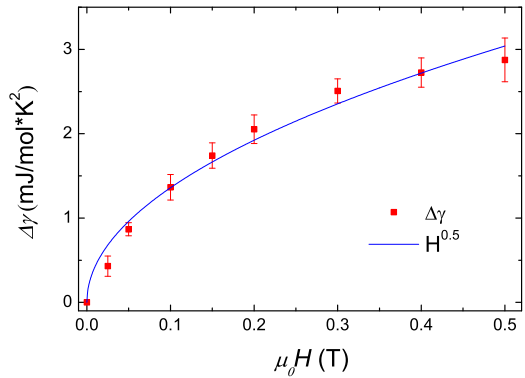}
\caption {(color online) The magnetic field dependence of the field induced electronic specific heat $\Delta\gamma_e$(H). The nonlinear field dependence is very clear. The blue line is a fit to the Volovik relation $\Delta \gamma_e = A\sqrt{H}$.} \label{fig4}
\end{figure}
To make the conclusion more convincing, we also derived the magnetic field dependence of $\gamma_e$. The $\gamma_e(H)$ is obtained by extending the low temperature specific heat to 0K at different magnetic fields. Fig.~4 shows the relation between $\Delta\gamma_e=C(H)/T-C(0)/T=\gamma_e(H)-\gamma_e(0T)$ and magnetic field at T = 0K. It is clear that $\Delta \gamma_e(H)$ is close to the Volovik relation, namely $\Delta \gamma_e(H)\propto\sqrt{H}$, which indicates again the presence of gap nodes. Actually the observations of $\gamma_e(T)\propto T$ and $\Delta \gamma_e(H)\propto \sqrt{H}$ present self-consistent evidence for the existence of line nodes, which gives a further enforcement on the gap structure comparing with the earlier thermal transport measurement\cite{FeSHT}. We should emphasize, however, that we could not distinguish whether the gap nodes are induced by the sign change of the gap, like in a d-wave case, or it is induced by the accidental nodes as theoretically predicted for the FeAs- and FeSe- based superconductors\cite{hirschfeldReview}. This needs to be resolved by more investigations.

\begin{table}
\caption{Fitting parameters with different models for FeS }
\begin{tabular}
{ccccccc}\hline \hline
model & $\Delta_1$(meV) & p   & $\Delta_2$(meV) & 1-p \\
\hline
$s-wave$  & 0.5  & 100\%    &  -     &  - \\
$d-wave$  & 0.6  & 100\%    &  -     &  - \\
$s-wave+d-wave$       & 0.6(s)  & 23\%     &  0.5(d)  & 77\% \\
$two \qquad d-wave$       & 0.6  & 91\%     &  1.0  & 9\% \\
 \hline \hline
\end{tabular}
\label{tab.1}
\end{table}

\begin{figure}
\includegraphics[width=8cm]{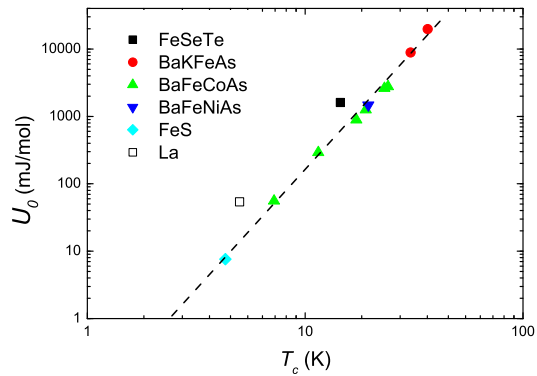}
\caption {(color online) Correlations between the condensation
energy and T$_c$ in FeS and several iron based systems. The condensation energy is calculated through integrating the entropy in the superconducting state.} \label{fig5}
\end{figure}

In iron-based superconductors a power law correlation $\Delta C|_{T_c}\propto T_c^3$, named as the BNC relation, was discovered for many iron based superconductors\cite{BNC}. In addition, another relation $U_0\propto T_c^n$ between condensation energy $U_0$ and superconducting transition temperature $T_c$ was also reported\cite{condensation} with $n\approx 3.5\pm0.5$, which was explained by the quantum critical phenomenon\cite{Zaanen}. So this inspires us to investigate whether the new superconductor FeS is also satisfied with these relations. Interestingly we find that the power law relation $\Delta C|_{T_c}\propto T_c^3$ is roughly obeyed if we put the data of FeS on the general plot. From the electronic specific heat data of FeS, we calculate the condensation energy through integrating the entropy of superconducting state and normal state by the formula below

\begin{eqnarray}
U_0=\int_{0}^{T_c}(S_n(T)-S_s(T))dT \qquad \qquad \qquad\nonumber\\
\nonumber\\
=\int_{0}^{T_c}dT\int_{0}^T(C_n(T')-C_s(T'))/T'dT'.
\end{eqnarray}

The calculated condensation energy of FeS is about 7.6mJ/mol. In Fig.~5 we present the condensation energy of FeS and other iron-based superconductors measured previously in our group\cite{condensation} together with that of element La. Intriguingly, the data of FeS falls into the power law relation perfectly. This may hint on that the FeS superconductor has similar mechanism as other FeAs- or FeSe-based superconductors. However, both the specific heat anomaly and $H_{c2}(0)$ in FeS are much lower than that in FeSe crystals\cite{FeSeSH}. We must point out that either the BNC scaling\cite{BNC} or the relation between the condensation energy and $T_c$\cite{condensation} do not apply to the KFe$_2$As$_2$ since it has a very large specific heat anomaly, which may suggest a strong renormalization of electron mass in that system\cite{Hardy}. In order to have comparison with a conventional superconductor, we also calculate the condensation energy of element La with T$_c$ = 5K which is similar to that of FeS\cite{Lacondensationenergy}. Although the T$_c$ of La is slightly higher than FeS, the condensation energy of La is about 54mJ/mol, which is almost 710\% of FeS. Therefore the data of La element deviates clearly from the power law that is satisfied by many iron based superconductors. This may suggest the different mechanism of the iron based superconductors and the conventional metal superconductors.

\section{\label{sec:level2}SUMMARY}

In summary, we have successfully grown the crystals of the newly discovered superconductor $\beta$-FeS and measured the low temperature specific heat. The electronic specific heat was obtained by removing the phonon contribution. A linear relation between the electronic specific heat and temperature was found in the low temperature region. By fitting to the theoretical models with different gap structures, we find that the model with nodal gap(s) fits the data best. The magnetic field induced specific heat shows the consistency with the Volovik relation $\Delta \gamma_e \propto\sqrt{H}$. All these suggest the existence of nodes on the gap structure. Our results will stimulate the investigations on the superconductor $\beta$-FeS and helps to resolve the intimate superconductivity mechanism in all iron based superconductors.

\begin{acknowledgments}
We appreciate the discussions with S. Y. Li. This work is supported by the NSF of
China, the Ministry of Science and Technology of China (973 projects: 2011CBA00102, 2012CB821403,
2010CB923002), and PAPD.
\end{acknowledgments}

$^{\star}$ hhwen@nju.edu.cn

\end{document}